\documentclass[Crown,sagev,times]{sagej}

\usepackage{moreverb,url}

\usepackage[colorlinks,bookmarksopen,bookmarksnumbered,citecolor=red,urlcolor=red]{hyperref}

\usepackage{bbold}
\usepackage{subcaption}
\usepackage{multirow}
\usepackage{graphicx}
\usepackage{booktabs}
\usepackage{mathtools }
\usepackage[ruled,vlined]{algorithm2e}
\mathtoolsset{showonlyrefs=true}
\usepackage[section]{placeins}

\newcommand{\Tau}{\varepsilon}
\usepackage{pgfplots}
\bibpunct{(}{)}{;}{a}{,}{,}  

\newtheorem{corollary}{Corollary}[section]

\newtheorem{proposition}{Proposition}[section]

\newcommand\BibTeX{{\rmfamily B\kern-.05em \textsc{i\kern-.025em b}\kern-.08em
T\kern-.1667em\lower.7ex\hbox{E}\kern-.125emX}}

\begin{document}

\runninghead{}

\title{A note on a resampling procedure for estimating the density at a given quantile}

\author{Beatriz Farah\affilnum{1,2}, Aurélien Latouche\affilnum{1,4}, Olivier Bouaziz\affilnum{2,3}}

\affiliation{\affilnum{1}Institut Curie, INSERM U1331, Mines Paris Tech, Paris/ Saint-Cloud, France \\
 \affilnum{2}Université Paris Cité, CNRS, MAP5, F-75006 Paris, France \\
 \affilnum{3}Univ. Lille, CNRS, UMR 8524 - Laboratoire Paul Painlevé, F-59000 Lille, France \\
 \affilnum{4}Conservatoire National des Arts et Métiers, Paris, France 
 }


\corrauth{Beatriz Farah, Institut Curie, INSERM U1331, Mines Paris Tech, Paris/ Saint-Cloud, France}

\email{beatriz.farah@math.cnrs.fr \\
}

\begin{abstract}
  In this paper we refine the procedure proposed by \cite{lin2015conditional} to estimate the density at a given quantile based on a resampling method. 
  The approach consists on generating multiple samples of the zero-mean Gaussian variable from which a least square estimator is constructed. 
  The main advantage of the proposed method is that it provides an estimation directly at the quantile of interest, thus achieving the $n^{-1/2}$ parametric rate of convergence.
  In this study, we investigate the critical role of the variance of the sampled Gaussians on the accuracy of the estimation.
  We provide theoretical guarantees on this variance that ensure the consistency of the estimator, and we propose a grid-search algorithm for automatic variance selection in practical applications.
  We demonstrate the performance of the proposed estimator in simulations and compare the results with those obtained using kernel density estimator.
\end{abstract}


\maketitle

\section{Introduction}
Quantiles have become a central tool in various statistical applications, namely econometrics and biostatistics, due to their robustness to outliers and their straightforward interpretation. 
This measure is particularly valuable in fields such as survival analysis and medical research, since quantiles of survival times allow the benefit, if any, of one treatment over another to be easily communicated in the timescale. 

In those applications, the density at a fixed quantile plays an important role in several inference procedures, including the asymptotic distribution of sample and regression quantiles, the construction of confidence intervals, hypothesis testing for equality of quantiles or regression parameters, and the estimation of standard errors for quantile estimators. This quantity, referred to as the \textit{quantile density function} in~\cite{parzen1979nonparametric}, is often seen as a nuisance parameter whose estimation is crucial in quantile estimation problems. This problem is not new, as described in the work from~\cite{tukey1965part} who refers to this quantity as the \textit{sparsity function}. A high density at the quantile will correspond to a low sparsity of the observations, which amounts to a better precision in the quantile estimation, and reciprocally.  
%
%
%
As an illustration, the density at the quantile is essential for determining the asymptotic precision of quantiles in two sample and regression problems~\citep{koenker2001quantile}. For the latter case, 
building confidence intervals, Wald, likelihood and score tests for regression parameters requires the estimation of this quantity~\citep[see][]{koenker1978regression, koenker1994confidence, koenker1982tests}. 
Estimation of the density at the quantile also arises in reliability analysis, where the focus is on the hazard quantile function, the mean residual quantile function, or on the total time on test transforms \citep{nair2013quantile}. 

A number of methods have been proposed for estimating the density at the quantile based on regression quantiles, particularly in the context of location and linear regression models. Early works considered order statistics based estimators involving a bandwidth parameter that needs to be tuned~\citep[see][]{siddiqui1960distribution, bloch1968simple, bofingeb1975estimation}. In the one-sample case, those approaches were generalized using kernel density estimators in~\cite{yang1985smooth, falk1986estimation, zelterman1990smooth}. In the presence of censoring, kernel density estimator of the density can also be constructed as an Inverse-Probability-of-Censoring-Weighting (IPCW) estimator using the Kaplan-Meier estimator of the censoring distribution~\citep[see][]{foldes1981strong, kosorok1999two}. 
%
However, a drawback of these approaches is that they require the estimation of the density at all points of the function which results in a slow rate of convergence that depends on the regularity of the density. 
More precisely, for a density $\beta$-times differentiable, the rate of convergence is of order $n^{-\beta/(2\beta+1)}$ for an optimal bandwidth of order $n^{-1/(2\beta+1)}$. 
Moreover, this method relies on the specification of an unknown bandwidth parameter, which has an impact on the estimator's performance \citep{heidenreich2013bandwidth}.

A different approach, based on resampling, has been proposed in~\cite{lin2015conditional}. 
The method works for censored data and consists on generating multiple realizations of zero-mean Gaussian variables with variance equal to one. The density at the quantile is then obtained by least square estimation. 
Since this procedure estimates the density at a single point, it achieves the $n^{-1/2}$ parametric rate of convergence, which is faster than kernel density estimation.
Under the framework of~\cite{lin2015conditional}, the key contribution of our work is a rigorous analysis of the impact of the variance of the sampled Gaussians on the accuracy of the least square estimator. In particular, we show that the estimator converges in probability to the target density for values of the standard deviation that lie inside an interval of the type $(C_1 n^\alpha, C_2 n^\alpha)$, with $C_1$, $C_2\in\mathbb R$ and $\alpha \in (-1/2, 1/2)$. We further study the Mean Squared Error (MSE) of this estimator empirically, and observe that there exists an interval of values for the standard deviation that gives a precise estimation of the density at the quantile. Moreover, in agreement with our theoretical results, this interval is shown to increase  with the sample size. 
Finally, for practical applications of our method, we also propose a grid-search algorithm in order to automatically select the variance of the sampled Gaussians. 

This paper is organized as follows. 
First, we review in Section \ref{sec:methods} the resampling procedure inspired by \citet{lin2015conditional}.
Next, we establish in Section \ref{sec:theory_variance} the theoretical guarantees for the consistency of the estimator.
From this result, we derive the interval of values of the standard deviation, that depends on sample size, for which the resampled estimator is guaranteed to converge to the target density.
In Section~\ref{sec:variance_selection}, the MSE is studied through empirical experiments and a grid-search algorithm for automatic variance selection is proposed for practical applications. 
Simulation studies are reported in Section \ref{sec:simulations} to assess the performance of the proposed estimator constructed using our automative variance selection rule and results are compared to the ones obtained by kernel density estimator. We conclude with a discussion in Section \ref{sec:discussion}.

\section{Methods}
\label{sec:methods}
In this section we present preliminary theoretical results that justify the validity of the proposed resampling procedure.
This is followed by a step-by-step description of the method. For more generality, we place ourselves in the context of right-censored data, but all our results are also valid when all events are observed. This is a direct consequence of the Kaplan-Meier estimator that reduces to the empirical distribution function in the absence of censoring.
\subsection{Preliminaries}
\label{sec:preliminaries}
Let $\tilde{T}$ be the event of interest, $C$ be the censoring variable, and assume that those variables are independent. We observe a sample $(T_i,\Delta_i)$, $i=1,\ldots,n$, where  $T_i = \min(\tilde{T}_i, C_i)$ and $\Delta_i = \mathbb{1}_{\tilde{T}_i \leq C_i}$. 
Let $p \in (0,1)$, the quantile of order $p$ is defined as $q = \inf \{t: F(t) \geq p \}$. Its estimator is $\hat q = \inf \{t: \hat F(t) \geq p \}$ where $1-\hat{F}$ is the usual Kaplan-Meier estimator of $1-F$. 
We assume $\tilde T$ to be continuous with $f$ its density function. Our goal is to estimate $f(q)$, the density evaluated at the quantile, which is assumed to be positive. We also assume $f$ to be differentiable and bounded in a neighborhood of $q$.  
Denoting the survival function of the observed times as $H(t)=\mathbb P[T\geq t]$, we suppose there exists $\tau > 0$ such that $H(q + \tau) > 0$.
This condition is needed to ensure sufficient follow-up in order to be able to estimate the quantile of interest. 
We have the following theoretical result.
\begin{proposition}
    Assume that the previously outlined conditions hold and let $\tilde{q} = \hat q + \Tau / \sqrt{n}$, where $\Tau\sim\mathcal N(0,\sigma^2)$. We have:
    \begin{align}
        \sqrt{n} (\hat q - q) & = -\frac{\sqrt{n}(\hat F(q) -p ) }{f(q)} + o_p(1),\label{eq:lemma_kosorok}\\
      \sqrt{n}(\tilde{q} - q) & = -\dfrac{\sqrt{n} (\hat{F}(q) - \hat{F}(\tilde{q} )) }{f(q)} + o_p(1). 
    \label{eq:lemma_f_tilde}
    \end{align}
\label{theorem1} 
\end{proposition}
The result presented in Equation~\eqref{eq:lemma_kosorok} is well-known and its proof is available in the literature (see for example~\cite{kosorok1999two}).
The proof for Equation~\eqref{eq:lemma_f_tilde} is detailed in the Supplemental material.

\subsection{Resampling procedure}
\label{sec:resampling}
Assuming the conditions outlined in Proposition~\ref{sec:preliminaries} hold, it follows by computing Equation~\eqref{eq:lemma_kosorok} minus Equation~\eqref{eq:lemma_f_tilde}:
\begin{equation}
\begin{aligned}
         \sqrt{n} ( \hat{F}(\tilde{q}) - p)&=  f(q) \Tau + o_p(1). 
\end{aligned}
\end{equation}
This result suggests using the following resampling procedure to estimate $f(q)$ as introduced in \cite{lin2015conditional}.
\renewcommand{\labelenumi}{Step \arabic{enumi}:}
\begin{enumerate}
    \item Generate B realizations of i.i.d Gaussian variables $\Tau_b \sim \mathcal{N}(0, \sigma^2)$, $b = 1,...,B$.
    \item Compute $\hat{Y}_b=\sqrt{n}\left( \hat{F}\left(\hat q + \Tau_b/\sqrt{n}\right) - p \right), b = 1,..., B$. Then the least squares estimate of $f(q)$ is $\widehat{f(q)}=(X^TX)^{-1}X \hat{Y}$, where $X= (\Tau_1,..., \Tau_B)^T$, $\hat{Y} = (\hat{Y}_1,..., \hat{Y}_B)^T$.
\end{enumerate}
\renewcommand{\labelenumi}{\arabic{enumi}.}
In practice, $B$ should be chosen as large as possible, such as $B=10^5$ for instance. In~\cite{lin2015conditional} the variance of the sampled Gaussians $\sigma^2$ is fixed and equal to 1.
However, in the remainder of this paper we show that the variance of the Gaussians plays an important role and should be carefully specified, as it impacts the quality of the estimation, especially for small sample sizes.
Specifically, in the next section we present a novel result for the consistency of the estimator, which highlights the influence of $\sigma^2$ on the convergence of the estimator to the target density.
We then provide experimental evidence to illustrate the impact of the variance on the accuracy of the estimation and, based on those findings, we propose an automatic method to select this variance.

\section{Theoretical results}
\label{sec:theory_variance}
In this section we present the formal mathematical development for the consistency of the estimator of $f(q)$. We now allow the variance of the Gaussian variables to depend on the sample size, that is $\Tau_b\sim\mathcal N(0,\sigma^2_n)$ and we aim at studying the effect of $\sigma_n$ on the consistency of $\widehat{f(q)}$.

Since we can control the size of $B$, we start by studying the least-square estimator as $B$ tends to infinity. We start by writing: 
\begin{align*}
\widehat{f(q)}=\left(\frac 1B\sum_{b=1}^B \Tau_b^2\right)^{-1}\frac 1B\sum_{b=1}^B \Tau_b \hat{Y}_b.
\end{align*}
From the law of large numbers and the continuous mapping theorem, the first term $(\sum_{b} \Tau_b^2/B)^{-1}$ converges in probability towards $1/\sigma_n^2$ as $B$ tends to infinity. The second term is studied in the next proposition.


\begin{proposition}
The quantity $\sum_{b}\Tau_b \hat{Y}_b/B$ converges in probability, as $B$ tends to infinity, towards $\mathbb E[\Tau_b \hat{Y}_b \mid T_{1:n}, \Delta_{1:n}]$, where the conditional expectation is taken with respect to $(T_1,\Delta_1),\ldots, (T_n,\Delta_n)$.
\label{prop:limit_B}
\end{proposition}
This result is proved in the Supplemental material. Now, using the independence between $\Tau_b$ and the observations, the limiting distribution in Proposition~\ref{prop:limit_B} can be expressed in the following way:
\begin{equation}
    \begin{aligned}
         \mathbb E[\Tau_b \hat{Y}_b \mid T_{1:n}, \Delta_{1:n}] &= \sqrt{n} \int_{-\infty}^{+\infty} u \left(\hat{F}\left( \hat{q} + \frac{u}{\sqrt{n}} \right) - \hat{F}(\hat{q}) \right) \varphi_{\sigma_n}(u) \, du,
    \end{aligned}
    \label{eq:expect_cond}
\end{equation}
where $\varphi_{\sigma_n}$ represents the probability distribution function of the centered Gaussian variable with variance $\sigma_n^2$. Then, we write:
\begin{align}
\sqrt{n} \left( \hat{F}\left( \hat{q} + \frac{u}{\sqrt{n}} \right) - \hat{F}(\hat{q}) \right) 
&= \sqrt{n} \left( \hat{F}\left( \hat{q} + \frac{u}{\sqrt{n}} \right) - F\left( \hat{q} + \frac{u}{\sqrt{n}} \right) \right) \nonumber \\
&\quad + \sqrt{n} \left( F(\hat{q}) - \hat{F}(\hat{q}) \right) 
+ \sqrt{n} \left( F\left( \hat{q} + \frac{u}{\sqrt{n}} \right) - F(\hat{q}) \right).
\end{align}
The term $ \sqrt{n} \left( F(\hat{q}) - \hat{F}(\hat{q}) \right) $ does not depend on $u$ and therefore vanishes in Equation~\eqref{eq:expect_cond}. On the other hand, using a Taylor expansion, we have:
\begin{equation}\label{eq:TaylorF}
    \sqrt{n} \left( F\left( \hat{q} + \frac{u}{\sqrt{n}} \right) - F(\hat{q}) \right) = \left( f(\hat{q}) u + f'\left(\hat{q} + \frac{s u}{\sqrt{n}}\right) \frac{u^2}{\sqrt n} \right) \mathbb{1}_{\hat{q}+\frac{u}{\sqrt{n}}>0} -\sqrt{n} F(\hat{q}) \mathbb{1}_{\hat{q}+\frac{u}{\sqrt{n}} <0},
\end{equation}
where $s \in (0,1)$. As a result, we have proved that $\widehat{f(q)}$ converges in probability, as $B$ tends to infinity, towards
\begin{align}
\frac {\sqrt{n}}{\sigma_n^2} \int u \left( \hat{F}\left( \hat{q} + \frac{u}{\sqrt{n}} \right) - F\left( \hat{q} + \frac{u}{\sqrt{n}} \right) \right)\varphi_{\sigma_n}(u) \, du + \frac {f(\hat{q}) }{\sigma_n^2}\int_{-\hat{q}\sqrt n}^{+\infty} u^2 \varphi_{\sigma_n}(u) \, du\nonumber\\
\quad + \frac{1}{\sigma_n^2\sqrt n}\int_{-\hat{q}\sqrt n}^{+\infty}f'\left(\hat{q} + \frac{s u}{\sqrt{n}}\right) u^3\varphi_{\sigma_n}(u) \, du- \frac{\sqrt{n}}{\sigma_n^2} F(\hat{q}) \int_{-\infty}^{-\hat{q}\sqrt n} u \varphi_{\sigma_n}(u) \, du.\label{eq:final_exp}
\end{align}
From this formula, each term needs to be analyzed separately in order to obtain the final result, which is stated in the next proposition. 
In the proof, it is seen that the condition $\sigma_n/\sqrt n\to 0$, as $n$ tends to infinity, is necessary in particularly to obtain that the term $ f(\hat{q})\int u^2 \varphi_{\sigma_n}(u) \mathbb{1}_{\hat{q}+\frac{u}{\sqrt{n}}>0} \, du/\sigma_n^2$ tends to $f(q)$ in probability.  For all the other terms to tend to $0$, we further need the condition $\sigma_n\sqrt n\to +\infty$. This gives a precise rate of convergence for $\sigma_n$, as detailled in the corollary.

\begin{proposition}\label{prop:main}
Assume that the conditions presented in Section~\ref{sec:preliminaries} are satisfied. We have:
\begin{align*}
\left|\frac 1{\sigma_n^2} \mathbb E[\Tau_b \hat{Y}_b \mid T_{1:n}, \Delta_{1:n}] - f(q) \right|\leq R_{n,1}+R_{n,2}+R_{n,3}+R_{n,4},
\end{align*}
with $R_{n,1}=O_{\mathbb P}(n^{-1/2})$, $R_{n,2}=O_{\mathbb P}(\sigma_n/\sqrt n)$, $R_{n,3}=\sigma_n^{-1/2}O_{\mathbb P}(n^{-1/4})$, $R_{n,4}=\sigma_n^{1/2}O_{\mathbb P}(n^{-3/4})$ and where the $O_{\mathbb P}(\cdot)$ are expressed as $n\to\infty$ and $\sigma_n/\sqrt n\to 0$.
        \label{lemma_positive}
\end{proposition}
\begin{corollary}
Assume the standard deviation of the sampled Gaussians satisfies the following inequalities:
    \begin{equation}
        C_1 n^\alpha < \sigma_n < C_2 n^\alpha,
    \end{equation}
    where $\alpha \in (-1/2, 1/2)$, $C_1, C_2 \in \mathbb{R}$. Then, under the conditions presented in Section~\ref{sec:preliminaries}, $\mathbb E[\Tau_b \hat{Y}_b \mid T_{1:n}, \Delta_{1:n}]/\sigma_n^2$ converges towards $f(q)$ as $n$ tends to infinity.
\end{corollary}
The proof of Proposition~\ref{prop:main} can be found in the Supplemental material. Of note, in the proof, we need to handle the case when a sample Gaussian variable $\Tau_b$ is such that $\hat q+\Tau_b/\sqrt n<0$. This is done using tail properties of the Gaussian distribution. An alternative definition of $\widehat{f(q)}$ could be proposed where only the samples $\Tau_b$ such that $\hat q+\Tau_b/\sqrt n$ is positive, are kept. This would lead to a similar estimator and the results stated in Proposition~\ref{prop:main} would still be valid with the same rate of convergence for $\sigma_n$.

\section{Variance selection of the sampled Gaussians}
\label{sec:variance_selection}
In the previous section, we have obtained theoretical results on the convergence in probability of the density estimator at the quantile. These results show that the variance of the sampled Gaussians should depend on the sample size and that the interval of values for this variance increases with the sample size. 
We are now interested in studying the empirical behavior for finite sample size. 
We begin this section by illustrating the impact of $\sigma_n$ on the Mean Squared Error (MSE) of the density for different sample sizes.
Then, inspired by our observations, we propose an automatic algorithm for variance selection.

\subsection{Empirical impact of the variance on density estimation}
\label{sec:impact_mse}
\begin{figure}[ht]
\centering
\begin{subfigure}[b]{0.7\textwidth}
\centering
\begin{tikzpicture}
\begin{axis}[
    width=\textwidth,
    height=0.5\textwidth,
    xlabel={$\sigma_n$},
    ylabel={MSE},
    title={$n = 50$},
    grid=major,
    grid style={line width=0.2pt, draw=gray!30},
    thick,
    line width=1pt,
    mark=*,
    mark options={fill=blue},
    mark size=1.0pt,
]
\addplot table [x="x", y="y", col sep=space, header=true]{data_n50.txt};
\end{axis}
\end{tikzpicture}
\caption*{}
\end{subfigure}
\hfill
\begin{subfigure}[b]{0.7\textwidth}
\centering
\begin{tikzpicture}
\begin{axis}[
    width=\textwidth,
    height=0.5\textwidth,
    xlabel={$\sigma_n$},
    ylabel={MSE},
    title={$n = 200$},
    grid=major,
    grid style={line width=0.2pt, draw=gray!30},
    thick,
    line width=1pt,
    mark=*,
    mark options={fill=red},
    mark size=1.0pt,
]
\addplot table [x="x", y="y", col sep=space, header=true]{data_n200.txt};
\end{axis}
\end{tikzpicture}
\caption*{}
\end{subfigure}

\vspace{1em}

\begin{subfigure}[b]{0.7\textwidth}
\centering
\begin{tikzpicture}
\begin{axis}[
    width=\textwidth,
    height=0.5\textwidth,
    xlabel={$\sigma_n$},
    ylabel={MSE},
    title={$n = 1000$},
    grid=major,
    grid style={line width=0.2pt, draw=gray!30},
    thick,
    line width=1pt,
    mark=*,
    mark options={fill=green},
    mark size=1.0pt,
]
\addplot table [x="x", y="y", col sep=space, header=true]{data_n1000.txt};
\end{axis}
\end{tikzpicture}
\caption*{}
\end{subfigure}
\caption{Illustration of the MSE of the density of the exponential at the median as a function of the standard error $\sigma_n$ for different sample sizes $n$.}
\label{fig:mse}
\end{figure}
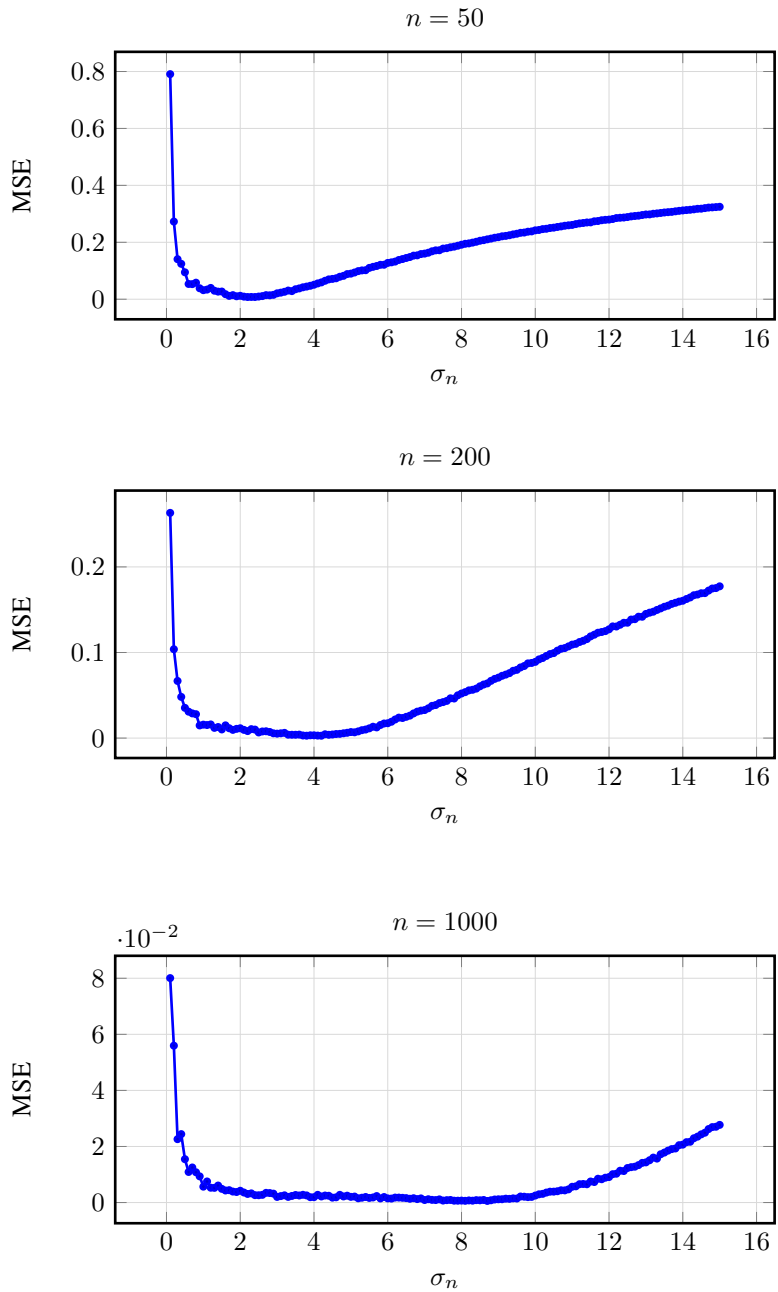

We consider a survival time $T$ that follows an exponential distribution with rate $0.12$ and whose density at the median is equal to $0.75$. The censoring $C$ is assumed to be independent of $T$ and to follow an exponential distribution with rate $0.12$. 
Using the method from~\cite{lin2015conditional}, the density at the median is estimated and the MSE is computed for a range of values of $\sigma_n$. 
For the resampling method, $B = 10^5$ zero-mean Gaussians are generated, with variance $\sigma_n^2$, $\sigma_n$ ranging from $0$ to $15$. The MSE is computed from $100$ Monte-Carlo replications  
and the results are seen in Figure \ref{fig:mse}, for sample sizes equal to $n=50$, $200$, $1000$.

For all sample sizes, we observe that the MSE has a pattern where it decreases until it reaches a plateau, where it remains low over a range of $\sigma_n$ values, before increasing again. 
As sample size increases, the region where the MSE is minimized becomes broader, which allows for greater flexibility when choosing the variance within a wider interval of values where the MSE is close to zero. This is in accordance with our theoretical results obtained in Section~\ref{sec:theory_variance}. In particular, the choice of $\sigma_n^2$ is particularly important for small sample sizes.

\FloatBarrier
\subsection{Grid-search algorithm for automatic variance selection}
\label{sec:grid_search}
In practice the true value of the density at the quantile is unknown, so we cannot directly compute the MSE to identify the plateau described earlier.
However, the presence of a plateau indicates that the estimation remains relatively stable over a certain interval of values of $\sigma_n$. 
Thus, our strategy consists in identifying the interval where the estimates exhibit minimal variation, which should correspond to a region with low MSE. 
To this end, we propose a grid-search algorithm to automatically identify such region from the pointwise estimates of the density at the quantile across a discrete grid of candidate $\sigma_n$ values.
This approach allows data-driven selection of $\sigma_n$ without requiring the knowledge of the underlying density and its derivatives.

Our method relies on analyzing the local behavior of a sequence of pointwise estimates of $f(q)$, computed over a grid of $\sigma_n$ values, using neighborhoods of fixed width $h$. This is a two-stage approach described in Algorithm~\ref{algo:plateau}. 
First, we search for local extrema by evaluating each point relative to its surrounding values within a symmetric window of size $2h+1$.
A point is identified as a candidate extremum if it attains either the maximum or minimum value within its local neighborhood. 
If multiple local extrema are found, we evaluate the local variation around each candidate by computing the sum of absolute differences between consecutive values in the neighborhood of width $h$ centered around each extremum, and we select the interval with the lowest local variation.
In the absence of local extrema found from step 1, we proceed to a second stage that analyzes the local variability of the vector of estimations.
We compute the first-order absolute differences between consecutive values of the vector of estimates and use a sliding window of width $h$ in order to identify regions of minimal cumulative variation. 
The region of size $2h+1$ with the lowest local variation is selected. Once the plateau is identified, through step 1 or step 2, we select a value for $\sigma_n$ inside the detected interval.

\begin{figure}[ht]
\centering
\begin{tikzpicture}
\begin{axis}[
    width=0.9\textwidth,
    height=0.5\textwidth,
    xlabel={$\sigma_n$},
    ylabel={$\widehat{f(F^{-1}(0.5))}$},
    grid=major,
    grid style={line width=0.2pt, draw=gray!20},
    thick,
    line width=1pt,
    mark=*,
    mark options={fill=blue},
    mark size=1.0pt,
    xmin=-0.5, xmax=10.5,
    ymin=0.2, ymax=0.9,
    legend style={at={(0.98,0.02)}, anchor=south east, font=\small},
]

\addplot[
    thick,
    line width=1pt,
    mark=*,
    mark options={fill=blue},
    mark size=1.0pt,
    color=blue,
    forget plot, 
] table [x="x", y="y", col sep=space, header=true]{data_f_sigmas.txt};


\addplot[
    black,
    thick,
    dashed,
    forget plot,
]
coordinates {(-0.5,0.75) (10.5,0.75)};

\addlegendimage{black, dashed, line width=1pt, no markers}
\addlegendentry{True $f(F^{-1}(0.5))$}

\addplot[
    red,
    thick,
    dotted,
    forget plot,
]
coordinates {(2.65,0.2) (2.65,0.9)};
\addlegendimage{red, dotted, line width=1pt, no markers}
\addlegendentry{Selected $\sigma_n$}

\end{axis}
\end{tikzpicture}
\caption{Example of estimation of the density at the median for the exponential distribution, for multiple $\sigma_n$.}
\label{fig:f_sigmas}
\end{figure}
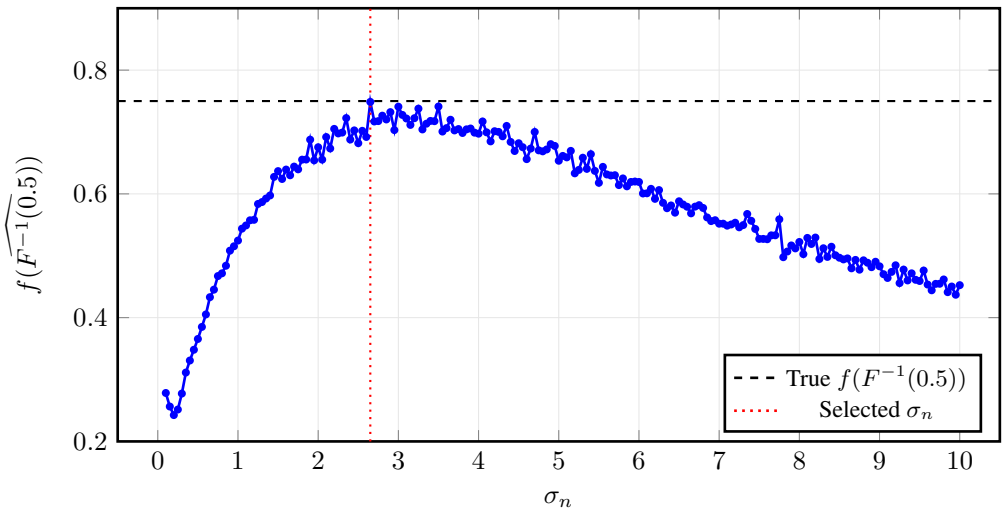
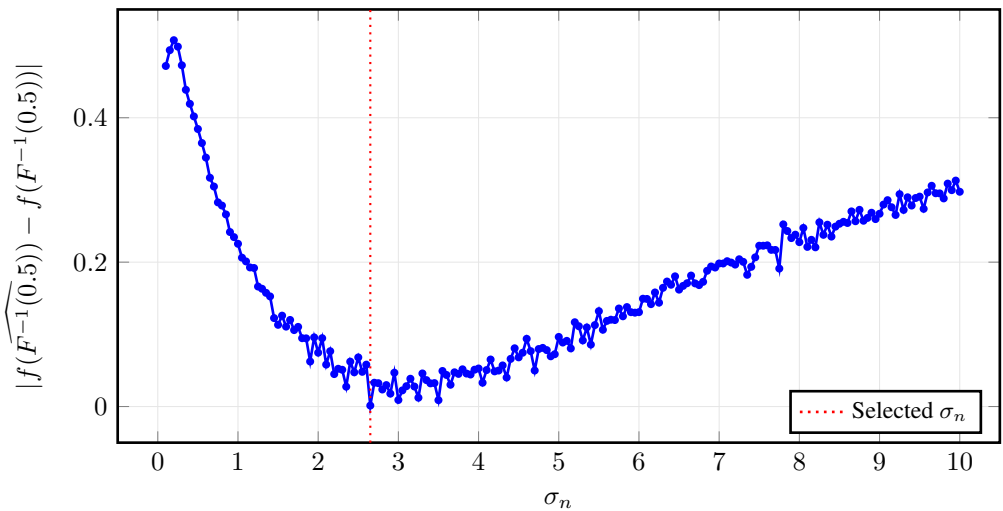
\begin{figure}[ht]
\centering
\begin{tikzpicture}
\begin{axis}[
    width=0.9\textwidth,
    height=0.5\textwidth,
    xlabel={$\sigma_n$},
    ylabel={$|\widehat{f(F^{-1}(0.5))} - f(F^{-1}(0.5))|$},
    grid=major,
    grid style={line width=0.2pt, draw=gray!20},
    thick,
    line width=1pt,
    mark=*,
    mark options={fill=blue},
    mark size=1.0pt,
    xmin=-0.5, xmax=10.5,
    ymin=-0.05, ymax=0.55,
    legend style={at={(0.98,0.02)}, anchor=south east, font=\small},
]

\addplot[
    thick,
    line width=1pt,
    mark=*,
    mark options={fill=blue},
    mark size=1.0pt,
    color=blue,
    forget plot, 
] table [x="x", y="y", col sep=space, header=true]{data_abs_diff.txt};



\addplot[
    red,
    thick,
    dotted,
    forget plot,
]
coordinates {(2.65,-0.05) (2.65,0.9)};
\addlegendimage{red, dotted, line width=1pt, no markers}
\addlegendentry{Selected $\sigma_n$}

\end{axis}
\end{tikzpicture}
\caption{Absolute value of the difference between the estimated $f(F^{-1}(0.5))$ and the true value.}
\label{fig:abs_diff}
\end{figure}

Our procedure is illustrated on a simulated data, with the survival and censoring times following an exponential distribution with rate $1.5$ and $0.12$, respectively. The true value of the density at the median is equal to $0.75$. 
The grid-search Algorithm~\ref{algo:plateau} is implemented with $\sigma_n$ ranging from $0$ to $10$, with increments of size $0.05$ and with neighborhoods of width $h = 20$, which results in intervals containing $41$ values. The procedure is displayed in Figure~\ref{fig:f_sigmas} on a simulated sample of size $n=200$. 
The true value of the density is represented by a dashed horizontal line. 
Our automatic grid-search method selects $\sigma_n = 2.65$, which is represented in the figure as a dotted vertical line. 
For this value, we obtain an estimation of the density at the median equal to $0.7486$. 
Finally, the absolute difference between the estimation and the true value of the density at the median is presented in Figure \ref{fig:abs_diff} for each $\sigma_n$. 
Similarly to the results presented in Figure~\ref{fig:mse} for $n = 200$, we observe a plateau of low MSE around a large interval of values for $\sigma_n$, roughly ranging from $2$ to $4$.

\begin{algorithm}
\caption{ \textbf{Grid-search algorithm for variance selection of the sampled Gaussians} }
\label{algo:plateau}
\SetKwInOut{Input}{Input}
\SetKwInOut{Output}{Output}

\SetKwInOut{Input}{Input}
\SetKwInOut{Output}{Output}

\Input{Grid of \(N\) values \(\sigma_1,\ldots, \sigma_N\). Vector of estimates \(\widehat{f(q)}_i\) at \(i=\sigma_1, \ldots, \sigma_N\). Neighborhood width \(h\).}
\Output{Selected value \(\sigma\)}
\hrule
\vspace{0.3em}
\textbf{Step 1: Identify local extrema with low variation}
\vspace{0.5em}
\hrule

Initialize empty list \texttt{candidate\_extrema}\;

\For{\(i \gets h+1\) \KwTo \(N - h\)}{
  Define neighborhood as \( [\widehat{f(q)}_{i-h}, \widehat{f(q)}_{i+h}] \)\;
  \If{\( \widehat{f(q)}_i = \max(\text{neighborhood}) \) \textbf{or} \( \widehat{f(q)}_i = \min(\text{neighborhood}) \)}{
    Append \(i\) to \texttt{candidate\_extrema}\;
  }
}

\If{\texttt{candidate\_extrema} is not empty}{
  Initialize \texttt{min\_variation} \(\gets \infty\), \texttt{best\_index} \(\gets \emptyset\)\;

  \For{ \textbf{each } \(c \in \texttt{candidate\_extrema}\)}{
    Compute local variation: \( V_c = \sum_{k = c - h}^{c + h - 1} |\widehat{f(q)}_{k+1} - \widehat{f(q)}_k| \)\;

    \If{\(V_c < \texttt{min\_variation}\)}{
      Update \texttt{min\_variation} \(\gets V_c\)\;
      Update \texttt{best\_index} \(\gets c\)\;
    }
  }

  Define plateau interval: \( [\sigma_{\texttt{best\_index} - h}, \sigma_{\texttt{best\_index} + h}] \)\;
  \Return midpoint of the plateau interval\;
}
\Else{
  Proceed to Step 2
}
\hrule
\vspace{0.3em}
\textbf{Step 2: Select interval with minimal local variation}
\vspace{0.5em}
\hrule

Compute absolute differences: \( d_i = |\widehat{f(q)}_{i+1} - \widehat{f(q)}_i| \) for \( i = 1 \) to \( N - 1 \)\;

Initialize \texttt{min\_variation} \(\gets \infty\), \texttt{best\_start} \(\gets 0\)\;

\For{\(j \gets 1\) \KwTo \(N - h\)}{
    Compute local variation in window: \( V_j = \sum_{k=j}^{j+h-1} d_k \)\;
    \If{\(V_j < \texttt{min\_variation}\)}{
      Update \texttt{min\_variation} \(\gets V_j\)\;
      Update \texttt{best\_start} \(\gets j\)\;
    }
  }

Define plateau interval \([\sigma_{\texttt{best\_start}}, \sigma_{\texttt{best\_start} + h}]\)\;

\Return midpoint of the plateau interval
\end{algorithm}


\section{Simulation study}
\label{sec:simulations}
In this section we conduct simulation experiments to compare the results for the estimation of the density at a quantile obtained with our resampling method and with kernel density estimation.
The variance for the sampled Gaussians is obtained through the grid-search Algorithm~\ref{algo:plateau}, as previously described, where we use a grid of $\sigma_n$ that ranges from $0$ to $10$, with increments of size $0.05$, and use neighborhoods of width $h=20$. 
For the  kernel density estimator, the bandwidth parameter is obtained through leave-one-out cross-validation and a Gaussian kernel is used. 
For all simulations we generate $B = 10^3$ zero-mean Gaussian variables and  $500$ Monte Carlo repetitions are implemented. 
We denote as LS our resampling procedure and as KDE the kernel density estimation. 
Details on the KDE method and our implementation can be found in the Supplemental material.

Two simulation scenarios are proposed.
In both scenarios, censoring times are modeled using exponential distributions, with varying rate parameters chosen to reach 10\%, 25\% and 40\%  of right-censored observations.

\begin{itemize}
    \item \underline{ Scenario 1: Exponential survival time with rate $1.5$.}
\end{itemize}

\begin{itemize}
    \item \underline{Scenario 2: Cauchy survival time with location and scale parameters equal to 0 and 1, respectively.}
\end{itemize}


\begin{table}[h!]
    \centering
    \begin{tabular}{ccccc}
        \hline
        \multicolumn{5}{c}{\rule{0pt}{4mm}\textbf{Sample Size: $n_i = 50$}} \\
        \hline
        \textbf{Censoring} & \textbf{Estimator} & \textbf{Bias} & \textbf{Variance} & \textbf{MSE} \\
        \hline
        40\% & LS  & -0.0090 & 0.0245 & 0.0245 \\
             & KDE & -0.1814 & 0.0029 & 0.0358 \\ \hline
        25\% & LS  & -0.0132 & 0.0201 & 0.0203 \\
             & KDE & -0.1834 & 0.0019 & 0.0356 \\ \hline
        10\% & LS  & -0.0170 & 0.0174 & 0.0177 \\ 
             & KDE & -0.1808 & 0.0015 & 0.0342 \\
        \hline
        \multicolumn{5}{c}{\rule{0pt}{4mm}\textbf{Sample Size: $n_i = 200$}} \\
        \hline
        \textbf{Censoring} & \textbf{Estimator} & \textbf{Bias} & \textbf{Variance} & \textbf{MSE} \\
        \hline
        40\% & LS  & 0.0538  & 0.0117 & 0.0146 \\
             & KDE & -0.1830 & 0.0006 & 0.0342 \\ \hline
        25\% & LS  & 0.0481  & 0.0088 & 0.0111 \\
             & KDE & -0.1831 & 0.0004 & 0.0340 \\ \hline
        10\% & LS  & 0.0470  & 0.0075 & 0.0097 \\
             & KDE & -0.1830 & 0.0003 & 0.0338 \\
        \hline
    \end{tabular}
    \caption{Results for Scenario 1.}
    \label{tab:stats_combined}
\end{table}

\begin{table}[h!]
    \centering
    \begin{tabular}{ccccc}
        \hline
        \multicolumn{5}{c}{\rule{0pt}{4mm}\textbf{Sample Size: $n = 50$}} \\
        \hline
        \textbf{Censoring} & \textbf{Estimator} & \textbf{Bias} & \textbf{Variance} & \textbf{MSE} \\
        \hline
        40\% & LS  & -0.0386 & 0.0046 & 0.0060 \\
             & KDE & -0.0106 & 0.0094 & 0.0095 \\ \hline
        25\% & LS  & -0.0090 & 0.0053 & 0.0054 \\
             & KDE &  0.0002 & 0.0083 & 0.0083 \\ \hline
        10\% & LS  & -0.0084 & 0.0050 & 0.0051 \\
             & KDE &  0.0050 & 0.0073 & 0.0073 \\
        \hline
        \multicolumn{5}{c}{\rule{0pt}{4mm}\textbf{Sample Size: $n = 200$}} \\
        \hline
        \textbf{Censoring} & \textbf{Estimator} & \textbf{Bias} & \textbf{Variance} & \textbf{MSE} \\
        \hline
        40\% & LS  & 0.0054 & 0.0031 & 0.0031 \\
             & KDE & -0.0490 & 0.0008 & 0.0032 \\ \hline
        25\% & LS  & 0.0066 & 0.0026 & 0.0027  \\
             & KDE & -0.0484 & 0.0005 & 0.0028 \\ \hline
        10\% & LS  & 0.0057 & 0.0026 & 0.0026 \\
             & KDE & -0.0484 & 0.0004 & 0.0028 \\
        \hline
    \end{tabular}
    \caption{Results for Scenario 2.}
    \label{tab:stats_combined_cauchy}
\end{table}




Table~\ref{tab:stats_combined} presents the results for Scenario 1. 
Across all sample sizes, the KDE method exhibits substantially higher bias compared to the LS method.
On the other hand, KDE has a smaller variance than LS, reduced by a factor of approximately 10. The MSE is smaller for LS at all sample sizes and censoring rates.
Table~\ref{tab:stats_combined_cauchy} shows the results for Scenario 2. 
In this case, for small sample sizes, KDE has a slightly lower bias than LS, while LS displays smaller variance.
As the sample size increases, LS outperforms KDE in terms of bias but with a bigger variance. As in Scenario 1, the LS method consistently achieves a lower MSE than KDE across all sample sizes and censoring levels. 
As expected, increased censoring leads to higher MSE for both methods, while larger sample sizes result in reduced MSE.

\section{Conclusion}
\label{sec:discussion}
In this paper we reviewed the method to estimate the density at a quantile through a resampling procedure inspired by \cite{lin2015conditional}.
Our work provided formal mathematical results for the consistency of the proposed estimator, which had not been previously established.
In particular, we emphasized the role of the variance of the sampled Gaussians on the accuracy of the estimation and showed that it should belong to an interval that depends on the sample size. As the sample size increases, the range of values for which the consistency of the density estimator is achieved increases. Those results are supported by our empirical experiments where we analyzed the effect of the variance on the MSE. Based on those findings, we then 
proposed a procedure for the automatic selection of the variance, which is extremely useful in practical applications. 
Finally, we compared simulation results obtained using our resampling technique as well as the kernel density estimation, and observed an improved performance in terms of MSE for our approach.

\section*{\normalsize Funding}
This work was supported by Ecole doctorale de Santé Publique, Université Paris-Saclay. 

\section*{\normalsize Disclosure statement}
The author(s) declared no potential conflicts of interest with respect to the research, authorship, and/or publication of this article.

\section*{\normalsize Data availability statement}
All data processing and statistical analysis were performed with the R statistical computing software version 4.3.2. 

\section*{\normalsize Supplemental material}
Supplemental material for this article is available online.

\bibliographystyle{plainnat}
\bibliography{biblio}

\end{document}



\title{Supplemental material: A note on a resampling procedure for estimating the density at a given quantile}






\maketitle

\tableofcontents

\newpage
\section{Proofs and Technical details}
In this section we provide the proofs of Proposition~3.1 and Proposition~3.2 of the main document. In the next subsection, we start by stating a result on the Kaplan-Meier estimator evaluated at the perturbed quantile value. 
We then recall some properties of the Gaussian distribution used in the proofs.
\subsection{Preliminaries}
The next lemma is a direct extension of Lemma 2 in~\cite{kosorok1999two}. Its proof is similar and therefore omitted.
\begin{lemma}
Under the assumptions presented in Section~2.1 of the main paper, for any interval $K_n=[-n^{\alpha/2},n^{\alpha/2}]$, with $\alpha<1$, we have
        \begin{equation*}\label{eq:claim}
            \sup_{u \in K_n} \left|\sqrt{n} \left(\hat{F}\left(\hat q + \dfrac{u}{\sqrt{n}}\right) - \hat{F}(\hat q) \right) - f(q)u\right| = o_p(1).
        \end{equation*}
        \label{lemma_aux1}
\end{lemma}
Standard results on the tail and on the expectation of the absolute value of a Gaussian variable are presented in the next two lemmas.
\begin{lemma}
Let $Z\sim \mathcal N(0,1)$, we have:
\begin{align*}
\mathbb P[Z\geq z] = O\left(\frac{1}{z}\exp\left(-\frac{z^2}{2}\right)\right),
\end{align*}
as $z$ tends to infinity.
\label{lemma_tail}
\end{lemma}

\begin{lemma}
Let $Z\sim \mathcal N(0,\sigma^2)$, we have:
\begin{align*}
\mathbb E[|Z|^a] = \frac{\sigma^a}{\sqrt \pi}2^{a/2}\Gamma\left(\frac{a+1}{2}\right),
\end{align*}
where $\Gamma$ is the standard gamma function.
\label{lemma_expectation}
\end{lemma}

\subsection{Proof of Proposition 2.1}

We only need to prove Equation~(2) of the main manuscript as Equation~(1) is a well know result (see for instance Lemma 1 in~\cite{kosorok1999two}). From the definition of $\tilde q$, we can directly write:
\begin{align*}
\sqrt n(\tilde q-q) & = \sqrt{n} (\hat q - q) +\Tau\\
& = -\frac{\sqrt{n}(\hat F(q) -\hat F(\tilde q))}{f(q)}-\frac{\hat F(\tilde q)-p}{f(q)}+\Tau + o_{\mathbb P}(1),
\end{align*}
where we used Equation~(1) of the main manuscript in the last equation. Now, from Lemma~\ref{lemma_aux1}, the term $|(\hat F(\tilde q)-p)/f(q)-\Tau|\mathbb 1_{\Tau\in K_n}$ converges to $0$ in probability as $n$ tends to infinity, where $K_n$ is defined as in Lemma~\ref{lemma_aux1}. On the other hand, since $|\hat F(\tilde q)-p|\leq 1$ and $f(q)>0$, the term $|(\hat F(\tilde q)-p)/f(q)-\Tau|\mathbb 1_{\Tau\in \bar K_n}$ is bounded and also converges to $0$ in probability, where $\bar K_n$ is the complementary set of $K_n$.

\subsection{Proof of Proposition 3.1}

%
%
%
%
%
    Let $\alpha > 0$, we have:
    \begin{equation}
        \begin{aligned}
            \alpha^2 \mathbb{1}_{|\frac{1}{B} \sum_{b = 1}^B \Tau_b \hat{Y}_b - \mathbb E(\Tau_b \hat{Y}_b \mid T_{1:n}, \Delta_{1:n})  | \geq \alpha} &\leq \left(\frac{1}{B} \sum_{b=1}^B \Tau_b \hat{Y}_b - \mathbb E(\Tau_b \hat{Y}_b \mid T_{1:n}, \Delta_{1:n}) \right)^2 \\
             \alpha^2 \mathbb E\left(\mathbb{1}_{|\frac{1}{B} \sum_{b = 1}^B \Tau_b \hat{Y}_b - E(\Tau_b \hat{Y}_b \mid T_{1:n}, \Delta_{1:n})  | \geq \alpha} \mid T_{1:n}, \Delta_{1:n} \right) &\leq \mathbb V\left(\frac{1}{B} \sum_{b=1}^B \Tau_b \hat{Y}_b  \mid T_{1:n}, \Delta_{1:n} \right) \\
            \mathbb P\left(\left|\frac{1}{B} \sum_{b = 1}^B \Tau_b \hat{Y}_b - E(\Tau_b \hat{Y}_b \mid T_{1:n}, \Delta_{1:n})\right| \geq \alpha \right)  &\leq \frac 1{\alpha^2 }\mathbb E\left(\mathbb V\left(\frac{1}{B} \sum_{b=1}^B \Tau_b \hat{Y}_b  \mid T_{1:n}, \Delta_{1:n} \right) \right).
            \end{aligned}
    \end{equation}
    Recall that $\hat{Y}_b=\sqrt n \left( \hat{F}\left(\hat q + \Tau_b/\sqrt{n}\right) - p \right)$ and use the independence between $\Tau_b$ and $(T_{1:n}, \Delta_{1:n} )$ to write
    \begin{align*}
        \mathbb V\left(\frac{1}{B} \sum_{b=1}^B \Tau_b \hat{Y}_b  \mid T_{1:n}, \Delta_{1:n} \right) & = \frac{n}{B}\mathbb V \left(\Tau_b \left( \hat{F}\left(\hat q + \Tau_b/\sqrt{n}\right) - p \right) \mid T_{1:n}, \Delta_{1:n} \right)\\
        & \leq  \frac{n}{B}\mathbb V \left(\Tau_b  \mid T_{1:n}, \Delta_{1:n} \right)= \frac{\sigma_n^2n}{B},
    \end{align*}
   where we used the fact that $ \hat{F} (\hat{q} + \Tau_b/\sqrt{n} ) - p) \leq 1$. Taking the limit as $B$ tends to infinity gives the final result.

%

\subsection{Proof of Proposition 3.2}

From our result, obtained in Section~3 of the main paper, we have $|E(\Tau_b \hat{Y}_b \mid T_{1:n}, \Delta_{1:n})/(\sigma_n)^{-2}-f(q)| \leq A_n+B_n+C_n+D_n$, with
\begin{align*}
A_n & =\left|\frac {f(\hat{q}) }{\sigma_n^2}\int_{-\hat{q}\sqrt n}^{+\infty} u^2 \varphi_{\sigma_n}(u) du-f(q)\right|,\\
B_n & = \frac{1}{\sigma_n^2\sqrt n}\left|\int_{-\hat{q}\sqrt n}^{+\infty}f'\left(\hat{q} + \frac{s u}{\sqrt{n}}\right) u^3\varphi_{\sigma_n}(u)du\right|,\\
C_n & = \frac{\sqrt{n}}{\sigma_n^2} F(\hat{q}) \left|\int_{-\infty}^{-\hat{q}\sqrt n} u \varphi_{\sigma_n}(u)du\right|,\\
D_n & = \frac {\sqrt{n}}{\sigma_n^2}\left|\int u \left( \hat{F}\left( \hat{q}+\frac{u}{\sqrt{n}} \right)- F\left( \hat{q} + \frac{u}{\sqrt{n}} \right)\right)\varphi_{\sigma_n}(u)du\right|.
\end{align*}

We start by studying the $A_n$ term. We have:
\begin{align*}
A_n & = \left|\frac {f(q) }{\sigma_n^2}\int_{-\hat{q}\sqrt n}^{+\infty} u^2 \varphi_{\sigma_n}(u) du-f(q)+\frac{f(\hat q)-f(q)}{\sigma_n^2}\int_{-\hat{q}\sqrt n}^{+\infty} u^2 \varphi_{\sigma_n}(u) du\right|\\
 & \leq f(q)\left|\frac { 1}{\sigma_n^2}\int_{-\hat{q}\sqrt n}^{+\infty} u^2 \varphi_{\sigma_n}(u) du-1\right| + \frac{\left|f(\hat q) -f(q)\right|}{\sigma_n^2} \int_{-\hat{q}\sqrt n}^{+\infty} u^2 \varphi_{\sigma_n}(u) du.
\end{align*}
From the mean value inequality theorem, for sufficiently large $n$, $|f(\hat q) -f(q)|\leq |\hat q -q | M_1$, for some constant $M_1>0$. Also, $\int u^2 \varphi_{\sigma_n}(u)\mathbb 1_{u\geq-\hat{q}\sqrt n} du/\sigma_n^2\leq 1$ and from Proposition 2.1, Equation~(1) from the main document, $|\hat q -q|=O_{\mathbb P}(n^{-1/2})$. This shows:
\begin{align*}
\frac{\left|f(\hat q) -f(q)\right|}{\sigma_n^2} \int_{-\hat{q}\sqrt n}^{+\infty} u^2 \varphi_{\sigma_n}(u) du = O_{\mathbb P}(n^{-1/2}).
\end{align*}
 On the other hand,
 \begin{align*}
\left|\frac { 1}{\sigma_n^2}\int_{-\hat{q}\sqrt n}^{+\infty} u^2 \varphi_{\sigma_n}(u) du-1\right| & = \frac { 1}{\sigma_n^2}\int_{-\infty}^{-\hat{q}\sqrt n} u^2 \varphi_{\sigma_n}(u) du,\\
& \leq \frac { 1}{\sigma_n^2} \left(\int u^4\varphi_{\sigma_n}(u) du\right)^{1/2}\left(\int_{-\infty}^{-\hat q\sqrt n}\varphi_{\sigma_n}(u) du\right)^{1/2}\!\!,
\end{align*}
from the Cauchy-Schwarz inequality. From Lemma~\ref{lemma_expectation}, the term $ (\int u^4\varphi_{\sigma_n}(u) du)^{1/2}/\sigma_n^2$ is bounded by a constant. Finally,
\begin{align*}
\int_{-\infty}^{-\hat q\sqrt n}\varphi_{\sigma_n}(u) du = O_{\mathbb P}\left(\frac{\sigma_n}{{\sqrt n}} \exp\left(-\frac{q^2n}{2\sigma_n^2}\right)\right), 
\end{align*}
as $\sigma_n/\sqrt n$ tends to $0$, from Lemma~\ref{lemma_tail} and the consistency of $\hat q$ towards $q$. This proves that 
\begin{align*}
A_n=O_{\mathbb P}(n^{-1/2}) + O_{\mathbb P}\left(\frac{\sqrt\sigma_n}{{n^{1/4}}} \exp\left(-\frac{q^2n}{4\sigma_n^2}\right)\right).
\end{align*}

For $B_n$, the quantity $|f'(\hat{q} + s u/\sqrt{n})|$ is bounded by a constant for $n$ large enough. Then
 \begin{align*}
\frac{1}{\sigma_n^2\sqrt n}\int_{-\hat{q}\sqrt n}^{+\infty} |u|^3 \varphi_{\sigma_n}(u) du & \leq \frac { 1}{\sigma_n^2\sqrt n} \left(\int u^6\varphi_{\sigma_n}(u) du\right)^{1/2}\left(\int_{-\hat{q}\sqrt n}^{+\infty}\varphi_{\sigma_n}(u) du\right)^{1/2}\!\!,
\end{align*}
from the Cauchy-Schwarz inequality. From Lemma~\ref{lemma_expectation}, the term $ (\int u^6\varphi_{\sigma_n}(u) du)^{1/2}/(\sigma_n^2\sqrt n)$ is bounded by $M_2 \sigma_n/\sqrt n$, for some constant $M_2>0$. Next, 
\begin{align*}
\int_{-\hat{q}\sqrt n}^{+\infty}\varphi_{\sigma_n}(u) du = \mathbb P[Z \geq -\hat{q}\sqrt n/\sigma_n],
\end{align*}
where $Z\sim\mathcal N(0,1)$. From the consistency of $\hat q$, this integral converges to $1$ in probability, as $\sigma_n/\sqrt n$ tends to $0$. This proves that:
\begin{align*}
B_n=O_{\mathbb P}\left(\frac{\sigma_n}{\sqrt n}\right).
\end{align*}

For the $C_n$ term, we use the Cauchy-Schwarz inequality and Lemma~\ref{lemma_expectation} as before to obtain
\begin{align*}
\frac{\sqrt{n}}{\sigma_n^2} F(\hat{q}) \left|\int_{-\infty}^{-\hat{q}\sqrt n} u \varphi_{\sigma_n}(u)du,\right| & \leq \frac{\sqrt{n}}{\sigma_n^2}  \left(\int u^2\varphi_{\sigma_n}(u) du\right)^{1/2}\left(\int_{-\infty}^{-\hat{q}\sqrt n}\varphi_{\sigma_n}(u) du\right)^{1/2}\!\!,\\
 &  \leq M_3\frac{\sqrt{n}}{\sigma_n} \left(\int_{-\infty}^{-\hat{q}\sqrt n}\varphi_{\sigma_n}(u) du\right)^{1/2}\!\!,
\end{align*}
for some $M_3>0$. The integral term is handled as before using Lemma~\ref{lemma_tail} and we finally obtain 
\begin{align*}
C_n=O_{\mathbb P}\left(\frac{n^{1/4}}{\sqrt\sigma_n} \exp\left(-\frac{q^2n}{4\sigma_n^2}\right)\right),
\end{align*}
as $\sigma_n/\sqrt n$ tends to $0$. 

We conclude with the $D_n$ term. First, introduce $N_i(t) = \mathbb 1_{T_i\leq t,\Delta_i=1} $ the observed counting process, $Y_i(t)=\mathbb 1_{T_i\geq t}$ the associated at-risk process and $\bar{N}(t)=\sum_i N_i(t)$, $\bar{Y}(t)=\sum_i Y_i(t)$. Let $\lambda$ and $\Lambda$ be the hazard and cumulative hazard rates, respectively. We have  $\bar{N}(t)-\int\bar Y(u)\lambda(u)\mathbb 1_{u\leq t}du=\bar M(t)$, with $\bar M(t)$ a martingale process with respect to the filtration $\mathcal F_t=\sigma(\bar N(u),\bar Y(u) : 0\leq u\leq t)$. Now write (see for instance~\cite{andersen2012statistical}):
\begin{align*}
& \sqrt n\left(\hat{F}\left( \hat{q}+\frac{u}{\sqrt{n}} \right)- F\left( \hat{q} + \frac{u}{\sqrt{n}}\right)\right)\\
& \quad = \left( 1 - F \left(\hat{q} + \frac{u}{\sqrt{n}} \right)\right) \sqrt{n}  \int_0^{\hat{q}+\frac{u}{\sqrt{n}}}\frac{1-\hat F(t-)}{1-F(t)} \frac{d\bar M(t)}{\bar{Y}(t)}\mathbb 1_{ \hat{q}+\frac{u}{\sqrt{n}}>0}.
\end{align*}
We can decompose the integral as:
\begin{align*}
\sqrt{n}  \int_0^{\hat{q}+\frac{u}{\sqrt{n}}}\frac{1-\hat F(t-)}{1-F(t)} \frac{d\bar M(t)}{\bar{Y}(t)} & = \sqrt{n}  \int_0^{\hat{q}+\frac{u}{\sqrt{n}}}\frac{d\bar M(t)}{\bar{Y}(t)}\\
&\quad -\sqrt{n}  \int_0^{\hat{q}+\frac{u}{\sqrt{n}}}\frac{\hat F(t-)-F(t)}{1-F(t)} \frac{d\bar M(t)}{\bar{Y}(t)}\cdot
\end{align*}
From standard results on martingales, the asymptotic normality of the Kaplan-Meier estimator and the consistency of $\hat q$, we see that the second term in the right hand side of the equality is of order $O_{\mathbb P}(n^{-1/2})$. The first term on the right-hand side of the equation can be decomposed as:
\begin{align*}
\sqrt{n}  \int_0^{\hat{q}+\frac{u}{\sqrt{n}}}\frac{d\bar M(t)}{\bar{Y}(t)} = \sqrt{n}\int_0^{\hat{q}}\frac{d\bar M(t)}{\bar{Y}(t)} + \sqrt n \left(\int_{\hat{q}}^{\hat{q}+\frac{u}{\sqrt{n}}}\frac{d\bar M(t)}{\bar{Y}(t)}\mathbb 1_{u>0}-\int_{\hat{q}+\frac{u}{\sqrt{n}}}^{\hat{q}}\frac{d\bar M(t)}{\bar{Y}(t)}\mathbb 1_{u<0}\right).
\end{align*}
Using a Taylor expansion for $F \left(\hat{q} + \frac{u}{\sqrt{n}} \right)$ and gathering all parts gives:
\begin{align}\label{eq:hard}
& \sqrt{n}\left( \hat{F}\left( \hat{q}+\frac{u}{\sqrt{n}} \right)- F\left( \hat{q} + \frac{u}{\sqrt{n}}\right)\right)\nonumber\\
& \quad = \left( 1 - F \left(\hat{q}\right) - \frac{u}{\sqrt{n}}f\left( \hat{q}+\frac{us}{\sqrt{n}} \right) \right)\left(\sqrt{n}\int_0^{\hat{q}}\frac{d\bar M(t)}{\bar{Y}(t)}+ \sqrt{n}  \int_{\hat{q}}^{\hat{q}+\frac{u}{\sqrt{n}}}\frac{d\bar M(t)}{\bar{Y}(t)}\mathbb 1_{u>0}\right.\nonumber\\
& \qquad -\left.\int_{\hat{q}+\frac{u}{\sqrt{n}}}^{\hat{q}}\frac{d\bar M(t)}{\bar{Y}(t)}\mathbb 1_{u<0} + O_{\mathbb P}(n^{-1/2})\right),
\end{align}
for $0<s<1$. Then
\begin{align}\label{eq:martingale}
\sqrt{n}  \int_{\hat{q}}^{\hat{q}+\frac{u}{\sqrt{n}}}\frac{d\bar M(t)}{\bar{Y}(t)}\mathbb 1_{u>0} &= \sqrt{n}  \int_{\hat{q}}^{\hat{q}+\frac{u}{\sqrt{n}}}\left(\frac{d\bar N(t)}{\bar{Y}(t)}-\lambda(t)dt\right)\mathbb 1_{u>0}\nonumber\\
& = \frac{\sqrt{n}}{n}\sum_{i=1}^n  \int_{\hat{q}}^{\hat{q}+\frac{u}{\sqrt{n}}}\left(\frac{dN_i(t)}{H(t)}-\lambda(t)dt\right)\mathbb 1_{u>0}\nonumber\\
&\quad -\frac{\sqrt{n}}{n}\sum_{i=1}^n  \int_{\hat{q}}^{\hat{q}+\frac{u}{\sqrt{n}}}\frac{\bar Y(t)/n-H(t)}{\bar Y(t)/n}\left(\frac{dN_i(t)}{H(t)}-\lambda(t)dt\right)\mathbb 1_{u>0}\nonumber\\
&\quad -\sqrt{n}  \int_{\hat{q}}^{\hat{q}+\frac{u}{\sqrt{n}}}\frac{\bar Y(t)/n-H(t)}{\bar Y(t)/n}\lambda(t)dt\mathbb 1_{u>0},
\end{align}
where $H(t)=\mathbb P[T\geq t]$. For the last term, note that, $\sup_t \sqrt n|\bar Y(t)/n-H(t)|=O_{\mathbb P}(1)$ where the supremum is taken for $t$ in a neighborhood of $q$, $\bar Y(t)/n$ converges towards $H(t)$ which is bounded from below in a neighborhood of $q$, and 
\begin{align}\label{eq:meanvalineq}
\int_{\hat{q}}^{\hat{q}+\frac{u}{\sqrt{n}}} \lambda(t)dt\mathbb 1_{u>0} = \left(\Lambda\left(\hat{q}+\frac{u}{\sqrt{n}}\right)-\Lambda(\hat{q})\right)\mathbb 1_{u>0}\leq\frac u{\sqrt{n}} M_4\mathbb 1_{u>0},
\end{align}
for some constant $M_4>0$, from the mean value inequality theorem. Here, we used the fact that $\lambda(t)=f(t)/S(t)$ which is bounded from above in a neighborhood of $q$. As a result
\begin{align*}
\sqrt{n}  \int_{\hat{q}}^{\hat{q}+\frac{u}{\sqrt{n}}}\frac{|\bar Y(t)/n-H(t)|}{\bar Y(t)/n}\lambda(t)dt\mathbb 1_{u>0} = u \mathbb 1_{u>0} O_{\mathbb P}(n^{-1/2}).
\end{align*}
We now study, for $u>0$, the empirical process 
\begin{align*}
S_n(v) = \sqrt{n}  \int_{v}^{v+\frac{u}{\sqrt n}}\left(\frac{d\bar N(t)}{H(t)}-\lambda(t)dt\right) = \sqrt n\left(\frac{1}{n}\sum_{i=1}^n \frac{\mathbb 1_{v\leq T_i\leq v+u/\sqrt n}\Delta_i}{H(T_i)}-\int_v^{v+\frac{u}{\sqrt n}}\lambda(t)dt\right),
\end{align*}
for $v$ in a neighborhood of $q$. From the independence censoring assumption, we directly see that, for $u>0$:
\begin{align*}
\mathbb E\left[\frac{\mathbb 1_{v\leq T_i\leq v+u/\sqrt n}\Delta_i}{H(T_i)}\right] = \int_v^{v+\frac{u}{\sqrt n}}\lambda(t)dt,
\end{align*}
such that $S_n$ is centered. Then, for $u>0$ 
\begin{align*}
V(v):=\mathbb V\left[\frac{\mathbb 1_{v\leq T_i\leq v+u/\sqrt n}\Delta_i}{H(T_i)}\right] & = \mathbb E\left[\frac{\mathbb 1_{v\leq T_i\leq v+u/\sqrt n}\Delta_i}{H^2(T_i)}\right] - \left(\int_v^{v+\frac{u}{\sqrt n}}\lambda(t)dt\right)^2\\
& = \int_v^{v+\frac{u}{\sqrt n}} \frac{\lambda(t)}{H(t)}dt - \left(\int_v^{v+\frac{u}{\sqrt n}}\lambda(t)dt\right)^2.
\end{align*}
 Since $1/H(\cdot)$ is bounded, the class of functions
 \begin{align*}
\left(\frac{\mathbb 1_{v\leq T_i\leq v+u/\sqrt n}\Delta_i}{H(T_i)}-\int_v^{v+\frac{u}{\sqrt n}}\lambda(t)dt\right)/\sqrt{V(v)}
\end{align*}
is Donsker (see~\cite{van1996weak}) and $S_n(v)/\sqrt{V(v)}$ converges towards a Brownian motion. 
As a result, $\sup_v |S_n(v)|/\sqrt{V(v)}=O_{\mathbb P}(1)$. We can further study $V(v)$ using the result from Equation~\eqref{eq:meanvalineq}:
\begin{align*}
\int_v^{v+\frac{u}{\sqrt n}}\lambda(t)dt = u\cdot O_{\mathbb P}(n^{-1/2}).
\end{align*}
The same argument applies to $\int\lambda(t)/H(t)\mathbb 1_{v\leq t \leq v+u/ \sqrt n}dt$ and therefore $V(v)= u\cdot O_{\mathbb P}(n^{-1/2})$. This proves that $S_n(\hat q) = \sqrt u \cdot O_{\mathbb P}(n^{-1/4})$. Going back to Equation~\eqref{eq:martingale}, we have that:
\begin{align*}
\frac{\sqrt{n}}{n}\sum_{i=1}^n  \int_{\hat{q}}^{\hat{q}+\frac{u}{\sqrt{n}}}\left(\frac{dN_i(t)}{H(t)}-\lambda(t)dt\right)\mathbb 1_{u>0} & = \sqrt u \mathbb 1_{u>0} \cdot O_{\mathbb P}(n^{-1/4})
\end{align*}
and
\begin{align*}
& \frac{\sqrt{n}}{n}\sum_{i=1}^n  \int_{\hat{q}}^{\hat{q}+\frac{u}{\sqrt{n}}}\frac{\bar Y(t)/n-H(t)}{\bar Y(t)/n}\left(\frac{dN_i(t)}{H(t)}-\lambda(t)dt\right)\mathbb 1_{u>0}\\
 & \quad \leq M_4 \sup |\bar Y(t)/n-H(t)|\cdot \left|\frac{\sqrt{n}}{n}\sum_{i=1}^n  \int_{\hat{q}}^{\hat{q}+\frac{u}{\sqrt{n}}}\left(\frac{dN_i(t)}{H(t)}-\lambda(t)dt\right)\right|\mathbb 1_{u>0} = \sqrt u\mathbb 1_{u>0} \cdot O_{\mathbb P}(n^{-3/4}),
\end{align*}
with $M_4>0$. In Equation~\eqref{eq:hard}, the same rate of convergence can be obtained for $\int \mathbb 1_{\hat{q}+u\sqrt{n}\leq t\leq \hat{q}}d\bar M(t)/\bar{Y}(t)\mathbb 1_{u<0}$ and $\sqrt{n}\int \mathbb 1_{t\leq \hat{q}}d\bar M(t)/\bar{Y}(t)=O_{\mathbb P}(1)$ from standard martingale theory (see~\cite{andersen2012statistical}). To conclude, gathering all the terms in Equation~\eqref{eq:hard}, we have shown that:
\begin{align*}
D_n & =  O_{\mathbb P}(n^{-1/2}) + \frac {1}{\sigma_n^{1/2}}O_{\mathbb P}(n^{-1/4})+ \frac {1}{\sigma_n^{1/2}}O_{\mathbb P}(n^{-3/4})+\sigma_n O_{\mathbb P}(n^{-1})+\sqrt{\sigma_n} O_{\mathbb P}(n^{-3/4})\\
& \quad+\sqrt{\sigma_n} O_{\mathbb P}(n^{-5/4})\\
& =O_{\mathbb P}(n^{-1/2}) + \frac {1}{\sigma_n^{1/2}}O_{\mathbb P}(n^{-1/4}) + O_{\mathbb P}(\sigma_n/\sqrt n) + \sqrt{\sigma_n} O_{\mathbb P}(n^{-3/4}).
\end{align*}

\section{Kernel density estimation}\label{sec:kde}
In the presence of censoring, a typical kernel estimator of the density can be constructed by estimating the censoring distribution from the Kaplan-Meier estimator. In \cite{foldes1981strong}, \cite{diehl1988kernel} the following estimator has been proposed:
\begin{align*}
\hat{f}_h(\hat{q}) = \frac{1}{nh} \sum_{i=1}^n \frac{\delta_i}{\hat S_{\text{cens}}(\hat{q}_i)} K\left( \frac{T_i - t}{h} \right),
\end{align*}
where $\hat S_{\text{cens}}$ is the Kaplan-Meier estimator of the censoring survival function and $K$ a kernel satisfying standard conditions. In order to compute this estimator, the kernel and the bandwidth must be chosen. In our implementation, a Gaussian kernel was taken and the bandwidth was obtained from cross-validation. For the choice of the bandwidth, the goal of the method is to try to minimize the Integrated Squared Error (ISE), defined, for $k=1, 2$, as: 
\begin{align*}
\text{ISE}(\hat f_h) & = \int \left(\hat f_h(\hat{q})-f_k(\hat{q})\right)^2dt\\
 & = \int \hat f^2_h(\hat{q})dt-2 \int \hat f_h(\hat{q})f_k(\hat{q})dt  + \int f_k^2(\hat{q})dt.
\end{align*}
In this expression the last term does not depend on $h$ and can be omitted. On the other hand, the term $\int f_h(\hat{q})f_k(\hat{q})dt$ is estimated by:
\begin{align*}
\hat J(h) = \frac{1}{n(n-1)h}\sum_{i\neq j}K\left(\frac{T_i-T_j}{h}\right)\frac{\delta_{ik}\delta_{jk}}{\hat S_{\text{cens}}(\hat{q}_i)\hat S_{\text{cens}}(\hat{q}_j)}\cdot
\end{align*}
Using the consistency of the Kaplan-Meier estimator, it can easily be shown that this estimator converges in probability, as $n$ tends to infinity, towards $\int \hat f_h(\hat{q})f_k(\hat{q})dt$ (see \cite{marron1987asymptotically}). In conclusion, our cross-validated estimator is defined as the Gaussian kernel estimator with bandwidth chosen as the minimizer of $\int \hat f^2_h(\hat{q})dt-2 \hat J(h)$, that is:
\begin{align*}
\hat h = \argmin_h \left\{ \int \hat f^2_h(\hat{q})dt- \frac{2}{n(n-1)h}\sum_{i\neq j}K\left(\frac{T_i-T_j}{h}\right)\frac{\delta_{ik}\delta_{jk}}{\hat S_{\text{cens}}(\hat{q}_i)\hat S_{\text{cens}}(\hat{q}_j)}\right\}\cdot
\end{align*}
We refer the reader to \cite{marron1987asymptotically} for more details about the cross-validation bandwidth selector and theoretical results regarding its validity.

\bibliographystyle{plainnat}
\bibliography{biblio_supp_mat}

%
%
%
%
%
%
%
%
%
%
%
%
%
%